\documentclass[showpacs]{revtex4}
\usepackage[dvips]{epsfig}

\newcommand{\be}{\begin{equation}}
\newcommand{\ee}{\end{equation}}
\newcommand{\clH}{{\cal H}}
\newcommand{\clu}{{\cal U}}
\newcommand{\clV}{\mbox{\large ${\cal V}$}}
\newcommand{\clv}{{\cal V}}
\newcommand{\clT}{{\cal T}}
\newcommand{\clE}{{\cal E}}

\newcommand{\bea}{\begin{eqnarray}}
\newcommand{\eea}{\end{eqnarray}}
\newcommand{\hh}{\tilde{h}}
\newcommand{\prt}{\partial}
\newcommand{\thh}{\tau_{\hbar}}
\newcommand{\rgl}{\rangle}
\newcommand{\lgl}{\langle}
\newcommand{\ep}{\epsilon}
\newcommand{\tr}{\mbox{tr}}
\begin{document}

\title{Semiclassical approximation for a nonlinear
oscillator with dissipation}
\author{A. Iomin }
\affiliation{Department of Physics, Technion, Haifa, 32000, Israel.  }

\begin{abstract}
An $S$--matrix approach is developed
for the chaotic dynamics of a nonlinear oscillator with dissipation.
The quantum--classical crossover is studied in the framework of the 
semiclassical expansion for the $S$--matrix. Analytical expressions for 
the braking time and the $S$--matrix are obtained.

\end{abstract}


\maketitle

We consider here a semiclassical dynamics of a nonlinear 
oscillator with dissipation. The main objective is to find the breaking 
time of the quantum--classical crossover for the dissipative system. In 
the absence of dissipation, 
the breaking time, namely, the Ehrenfest time has been found \cite{bz} to 
scale logarithmically with 
respect to the Planck constant $\hbar$: $\thh=(1/\Lambda)\ln(I_0/\hbar) 
$, where $I_0$ is a characteristic action, and $\Lambda$ is a Lyapunov 
exponent. It characterizes 
the exact classical--to--quantum correspondence between the Hamiltonian 
equation of motion and the Ehrenfest ones \cite{bz,zasl,biz81,bz82}. 
The renewed interest to this time is related to the extensive 
studies of the chaotic scattering in cavities \cite{silvestr}, the 
Loschmidt echo \cite{benenti} and observation of essential 
deviation from the logarithmic scaling for systems with phase space 
structures \cite{izpre01,izpre03}. The  nonlinear oscillator since the 
first observation of the Ehrenfest time \cite{bz} is 
explored to study  the quantum--classical correspondence 
\cite{zasl,biz81,bz82,sok,angelo}.
This time specifies a fast (exponential) growth of quantum corrections to 
the classical dynamics due to chaos.
In the presence of dissipation the breaking time differs from the $\thh$
since  the classical dissipation  changes  the local 
instability of trajectories. The breaking time $\thh^{(d)}$ for a 
dissipative web map has been obtained \cite{izpre03} by the 
$c$-number projection of the Heisenberg equations on the coherent states 
basis. At the time $\thh^{(d)}$, the  quantum corrections is of the order 
of 1 and  destroy the (semi)classical behavior of the system. The subject 
of quantum dissipative chaos grew out of the the pioneering work on 
dissipative quantum maps \cite{graham}, and  various aspects of the 
extensive studies on quantum dissipative 
chaos are reflected in the recent reviews \cite{haake} as well.

We show here in the framework of a $S$--matrix approach  for chaotic
dynamics of the nonlinear oscillator with a dissipation rate $\gamma$, 
that the breaking time $\thh^{(d)}$ depends essentially on the ration
between the dissipation rate $\gamma$ and  the local instability 
characteristics the Lyapunov exponent $\Lambda$.
The Hamiltonian of the system can be written in the following 
non-Hermitian form
\be\label{s4_1}
\clH=
\hbar(\Omega-i\gamma/2)a^{\dag}a+\hbar^2\mu(a^{\dag}a)^2-
\hbar\ep(a^{\dag}+a)\delta_T(t) \, .
\ee
Creation and annihilation operators has the commutation rule
$[a,a^{\dag}]=1$.
The complex frequency $\omega_{\gamma}=\Omega-i\gamma/2$ determines the 
effective
frequency $\omega=\left[\Omega^2+\gamma^2/4\right]^{1/2} $
in the presence of a finite width of the levels $\gamma/2$, and $\mu$ is 
the nonlinearity. The perturbation is a train of 
$\delta$-functions: $\delta_T(t)=\sum_{n=-\infty}^{\infty}\delta(t-nT)$, 
characterized by the amplitude $\epsilon$ and the period $T$.
Evolution of the wave function is governed by the quantum map
\be\label{s4_2}
\Psi(t+T)=\clu(T)\Psi(t) \, ,
\ee
where the evolution operator $ \clu(T)$ over the period $T$ 
corresponds to 
a free dissipative motion and then a kick. In our case it is a  
product of the unitary evolution operator $U(T)$ and the decay 
operator $B$
\be\label{s4_3}
\clu(T)\equiv\clu=UB=e^{i\ep(a^{\dag}+a)}e^{-i[\Omega 
Ta^{\dag}a+
\hh(a^{\dag}a)^2]}e^{-\gamma Ta^{\dag}a/2} \, .
\ee
Here the dimensionless semiclassical parameter $ \hh=\hbar\mu T $
is introduced. In what follows this parameter is small: $\hh\ll 1$.
To describe chaotic dynamics of the open system it is necessary  to
construct an $S$--matrix. To this end we close the system by means
of the complementary conditions with an incident wave $\phi_-(t)$ and an
outgoing wave $\phi_+(t)$. Therefore, the quantum map (\ref{s4_2}) takes
the following new form \cite{yan,livsic}
\be\label{s4_4}
{\Psi(t+T)\choose \phi_+(t)}=\clV {\Psi(t) \choose \phi_-(t)} =
\left(\begin{array}{cc}\clu &~~\clu W_1 \\W_2 &~~S_0\end{array}\right)
{\Psi(t) \choose \phi_-(t)} \, ,
\ee
where $\clv$ is unitary: $\clv^{\dag}\clv=\mbox{\bf\large 1}$.
Operators $W_1,W_2$ and $S_0$ are  determined from dissipation
by solving the Schr\"odinger equation
on the period $T$:
\be\label{s4_6}
\Psi(t+T)=\clu\Psi(t)-\frac{i}{\hbar}\int_{t+0}^{t+T+0}\clu(s-t)
\tilde{W}_1(s-T)\phi_-(s-T)ds \, .
\ee
Taking the closure of the system in the form
$\tilde{W}_1(t)=W_1\sum_{n=-\infty}^{\infty}\delta(t-nT)$, 
we obtain after the Fourier transform respect to time that
the quantum map (\ref{s4_4}) takes the following form
\bea\label{s4_7}
e^{iET}\psi(E)=\clu\psi(E)-\frac{i}{\hbar}\clu W_1\phi_-(E)
\nonumber \\
\phi_+(E)=W_2\psi(E)+S_0\phi_-(E).
\eea
Relation between the incident and outgoing waves is determined by
the following expression
$\phi_+(E)=S(E)\phi_-(E) $,
where $S(E)$ is called the $S$--matrix \cite{yan}. From this definition and 
Eq. (\ref{s4_7}) the $S$--matrix reads
\be\label{s4_9}
S(E)=S_0-\frac{i}{\hbar}W_2\frac{1}{e^{-iET}-\clu}\clu W_1 \, .
\ee
It is known \cite{mello} (see also \cite{yan}) that the matrix $\clV$ can 
be parameterized  as follows
\be\label{s4_10}
\clV=\left(\begin{array}{cc}U\sqrt{1-\clT\clT^+} &~~-U\clT \\
\clT^+&~~\sqrt{1-\clT^+\clT}\end{array}\right) \, ,
\ee
where $\clT=i\sqrt{1-B^2}$, and 
$ W_1=-i\hbar B^{-1}\clT,~~W_2=-\frac{i}{\hbar}W_1^{\dag}B
=\clT^{\dag},~~S_0=B $.
 After the parametrization (\ref{s4_10}), the $S$--matrix reads
\be\label{s4_11}
S(E)=B-\sqrt{1-B^2}\frac{1}{e^{-iET}-\clu}\clu\sqrt{(1-B^2)/B^2} \, .
\ee
Now, we consider the following  autocorrelation function
\be\label{s4_12}
R(\clE)=\overline {\tr\left(S^{\dag}(E+\clE/2T)S(E-\clE/2T)\right)}-
\overline{\tr\left(S^{\dag}(E)S(E)\right)} \, .
\ee
The over-line means an average over the quasienergy $ET$ from
$0$ to $2\pi$. It corresponds to the averaged cross--section 
\cite{blumel}. The treatment of this form is analytically tractable, and 
in what follows we perform the semiclassical analysis for the correlation 
function. After simple calculations, we obtain
\be\label{s4_13}
R(\clE)=\sum_{t}e^{-i\clE t} \tr\left((\clu^{\dag})^t\clu^t-
2(\clu^{\dag})^{t+1}\clu^{t+1}+(\clu^{\dag})^{t+2}\clu^{t+2}\right) \, .
\ee
The evolution operator $\clu(T)$ in the power $T$ could be formally 
considered  as the evolution operator for an arbitrary time $T$ 
corresponding to the power, namely $\clu^t(T)\equiv \clu(t)$.
 
For the trace we take the following integration over the coherent states
basis considered in the initial moment $t=0$:
$ \tr(\dots)=
\int\frac{d^2\alpha}{2\pi}\lgl\alpha|\dots|\alpha\rgl $.
The action of the evolution operator  
\be\label{s3_3}
\clu(t)=\widehat{\exp}\left\{-i\int_0^td\tau\left[\omega_{\gamma}
a^{\dag}a+\hbar\mu(a^{\dag}a)^2-
\ep\delta_T(\tau)(a^{\dag}+a)\right]\right\} \, ,
\ee
on the basis $|\alpha\rangle$ can be calculated analytically in the 
framework of the semiclassical approximation \cite{bz82,sok}.
Here $\widehat{\exp}$ means $T$--ordering. Applying the 
Stratonovich--Hubbard transformation \cite{habbard} under the 
$T$--ordering, one obtains
\be\label{s3_4}
\widehat{\exp}\left[-i\hbar\mu T\int_0^td\tau(a^{\dag}a)^2/T\right]=
\int\prod_{\tau}\frac{d\lambda(\tau)}{\sqrt{4\pi i\hh}}
e^{i\int_0^td\tau\lambda^2(\tau)/4\hh}\cdot
e^{-i\int_0^td\tau\lambda(\tau)a^{\dag}a} \, ,
\ee
where we use that $ \hh=\hbar\mu T $ and $ t/T\rightarrow t $ is a
number of kicks represented in the continuous form.
We take into account that the harmonic oscillator, acting on the coherent
state, changes its phase only, and the perturbation acts as
a shift operator. Therefore, the wave function in the moment $t$ has the 
form of the following functional integral
\bea\label{s3_5}
\Psi(t)=U(t)|\alpha\rgl=\int\prod_{\tau}\left(d\lambda(\tau)/
\sqrt{4\pi i\hh}\right)
\exp\left[i\int_0^td\tau\lambda^2(\tau)/4\hh\right] \nonumber \\
\times\exp\left[i\ep\int_0^td\tau\delta_1(\tau)(\alpha_{\lambda}^*(\tau)+
\alpha_{\lambda}(\tau))/2\right]
\exp\left[-(1-e^{-\gamma T})\int_0^td\tau\delta_1(\tau)
|\alpha_{\lambda}(\tau)|^2\right] |\alpha_{\lambda}(t)\rgl \, ,
\eea
where
\be\label{s3_6}
\alpha(t)=e^{-i\phi_{\lambda}(t)}a(t)=
e^{-i\phi_{\lambda}(t)}\left[\alpha+i\ep\int_0^td\tau\delta_1(\tau)
e^{i\phi_{\lambda}(\tau)}\right] \, ,
\ee
\be\label{s3_7}
\phi_{\lambda}(t)=\int_0^td\tau[\omega_{\gamma} T+\lambda(\tau)]=
\int_0^td\tau[\Omega T-i\gamma T+\lambda(\tau)] \, .
\ee
Denoting by 
$\beta_{\lambda}=-i\int_0^td\tau\delta_1(\tau)\alpha_{\lambda}(\tau)$,
we obtain the following expression for the trace
\bea\label{s4_14}
{\cal M}(t)=&
\int\frac{d^2\alpha}{2\pi}\lgl\alpha|\clu^{\dag}(t)\clu(t)|\alpha\rgl=
\int\frac{d^2\alpha}{2\pi}\int\prod_{\tau}
\frac{d\lambda_1(\tau)d\lambda_2(\tau)}{4\pi\hh}\exp\left[
\frac{i}{4\hh}\int_0^td\tau
\left(\lambda_1^2(\tau)-\lambda_2^2(\tau)\right)\right]
\nonumber \\
&\times\exp\left[i\mbox{Im}(\alpha_{\lambda_2}^*\alpha_{\lambda_1}-
\ep\beta_{\lambda_1}-\ep \beta_{\lambda_2}^*)-\frac{1}{2}
|\alpha_{\lambda_1}-\alpha_{\lambda_2}|^2\right] \nonumber \\
&\times \exp\left[-(1-e^{-\gamma T})\int_0^td\tau
g(\tau)(|\alpha_{\lambda_1}(\tau)|^2+
|\alpha_{\lambda_2}(\tau)|^2)/2\right].
\eea
The last exponential  can be evaluated by the following crude but 
reasonable approximation for $ t\gg 1/\gamma T$
\[
\int_0^td\tau g(\tau)(|\alpha_{\lambda_1}(\tau)|^2+
|\alpha_{\lambda_2}(\tau)|^2)/2\approx|\alpha|^2\sum_{\tau=0}^t
\exp(-\gamma T\tau)\approx\frac{|\alpha|^2}{1-e^{-\gamma T}} \, ,
\]
that essentially simplifies the follows semiclassical consideration.
In the limit $\hh\ll 1$, the expression for the trace 
$M(t)$ is strongly simplified and evaluated analytically.
Following \cite{sok}, we perform the linear transform
$\lambda_1=2\mu+\hh\nu/2,~~\lambda_2=2\mu-\hh\nu/2 $,
with the Jacobian equals to $2\hh$.
After the  variables change, we obtain from (\ref{s3_6}) and
(\ref{s3_7}) the following semiclassical expressions for the second 
exponential in (\ref{s4_14})
\be\label{s4_15}
\alpha_{\lambda_2}^*(t)\alpha_{\lambda_1}(t)-\ep\beta_{\lambda_1}-
\ep\beta_{\lambda_2}^*\approx-\int_0^td\tau(i\hh\nu(\tau)+\gamma T)
e^{-\gamma T\tau}|a(\tau)|^2 \, ,
\ee
and 
\be\label{s4_16}
\left|\alpha_{\lambda_1}-\alpha_{\lambda_2}\right|^2\approx
\left|\int_0^t d\tau\hh\nu(\tau)e^{-\gamma T\tau/2}a(\tau)\right|^2 \, .
\ee
Here $e^{-\gamma T\tau/2}a(\tau)$ is defined in Eq. (\ref{s3_6}) and 
(\ref{s3_7}) for $\nu\equiv 0$. Now we carry out integration over 
$\nu(\tau)$ and $\mu(\tau)$ in the classical limit, neglecting the term 
of the order of $\hh^2$ of Eq. (\ref{s4_16}) in the exponential.
Integral over $\nu$ in this case gives
$\prod_{\tau}2\pi\delta\left(\mu-\hh|e^{-\gamma 
T\tau/2}a(\tau)|^2\right)$,
and it leads to the exact integration  over $\mu$ as well.
Finely, we obtain for Eq. (\ref{s4_14})
\be\label{s4_17}
{\cal M}(t)\approx \int\frac{d^2\alpha}{2\pi}\exp\left[-|\alpha|^2-(\gamma
T/\hh)\int_0^t d\tau I_{cl}(\tau,\alpha,\alpha^*)\right] \, .
\ee
This expression could be evaluated by different approximations. For
example, we can take approximately in the following form
$ {\cal M}(t)\approx\exp\left[-\gamma T t\lgl I\rgl\right] $,
where $\lgl I\rgl$ is some classical action averaged over time with zero
initial conditions. It corresponds to a chaotic attractor with a finite
phase space volume for a specific choice of the parameters 
$\mu,\gamma,\ep$ and $T$. Inserting this into Eq. (\ref{s4_13}), 
we obtain a finite expression for the correlation function in the form
\be\label{s4_18}
R(\clE)=\frac{(1-e^{-\gamma T})^2}{1-\exp(-i\clE-\gamma T\lgl I\rgl)} \, .
\ee
It is interesting to admit that the autocorrelation function $R(\clE)$
related to an averaged cross section corresponds to the Ericson 
fluctuations (see {\em i.g.} \cite{blumel}).  Considering that $ \clE,
\gamma T\lgl I\rgl $ are small and taking into account only the first two
terms in the expansion of the exponential in the denominator in
(\ref{s4_18}), we obtain approximately, the following expression
\be\label{s4_19}
\left|R(\clE)\right|^2\propto \frac{1}{1+\big(\clE/{\cal G}\big)^2} \, ,
\ee
where ${\cal G}=\gamma T\lgl I\rgl$. This Lorentzian
corresponds to the Erickson distribution of fluctuations, see {\em e.g.}
\cite{blumel}.

The important point of the consideration is the neglecting of the term
$\hh^2\left|\int d\tau\nu(\tau)a(\tau)\right|^2$. 
Therefore, to specify the validity condition of the performed 
semiclassical approximation and the obtained result one needs to evaluate 
${\cal M}(t)$ taking into account the omitting term $O(\hh^2)$.
To this end, at performing the integration over $\nu(\tau)$, we use  the 
following auxiliary expression 
\be\label{auxil}
\exp\left[-\frac{\hh^2}{2}|
\int_0^td\tau\nu(\tau)e^{-\gamma T\tau/2}a(\tau)|^2\right]=
\frac{2}{\pi\hh}\int d^2\xi e^{-2|\xi|^2/\hh}
\exp\left[-i\frac{\sqrt{\hh}}{2}Re\xi^*
\int_0^td\tau\nu(\tau)e^{-\gamma T\tau/2}a(\tau)\right]
\ee
Substituting (\ref{s4_15}), (\ref{s4_16}) and (\ref{auxil}) into
(\ref{s4_14}), one obtains the following expression  for the trace 
\bea\label{s3_p1}
{\cal M}=\frac{2}{\hh\pi}\int d^2\xi e^{-2|\xi|^2/\hh}
\int\prod_{\tau}\frac{d\mu(\tau) d\nu(\tau)}{2\pi}
\exp\left[-i\int_0^t d\tau\nu(\tau)\mu(\tau)\right]
\nonumber \\
\exp\left[i\hh\int_0^t d\tau\nu(\tau)
(\big|e^{-\gamma T\tau/2}a(\tau)+\xi\big|^2-|\xi|^2)\right] \, . 
\eea
The functional integration over $\nu(\tau)$ is exact and gives the
$\delta$--function in $\mu$. Hence the integration over $\mu(\tau)$ is
also exact. After these integrations we obtain from (\ref{s3_p1})
\be\label{s3_p2}
{\cal M}=\frac{2}{\hh\pi}\int d^2\xi e^{-2|\xi|^2/\hh}
\exp\left[-|\alpha|^2-\gamma T\int_0^t d\tau
\bar{I}_{cl}(\alpha ,\alpha^*,\tau)\right] \, ,
\ee
where $\bar{I}_{cl}(\alpha,\alpha^*,\tau)\equiv I_{cl}(\Omega T-2|\xi|^2,
e^{-\gamma T\tau/2}a(\tau)+\xi,e^{-\gamma T\tau/2}a^*(\tau)+\xi^*)$, and 
$\alpha(\tau),~\alpha^*(\tau)$ are determined in Eqs. (\ref{s3_6}) and
(\ref{s3_7}). Expanding the last exponential in (\ref{s3_p2}) in the
Taylor series in $\xi$ and $\xi^*$, and taking into account that
$ (2/\pi\hh)\int d^2\xi e^{-2|\xi|^2/\hh}\xi^p\xi^{*q}=
\sqrt{(\hh/2)^{p+q}}\sqrt{p!q!}\delta_{p,q} $, 
we obtain the trace of Eq. (\ref{s3_p2}) in the form of the expansion
in the semiclassical parameter $\hh$
\be\label{s3_15}
{\cal M}=\sum_{n,l}\frac{(n+l)!}{(n!)^2l!}
\frac{\prt^{2n}}{\prt\xi^n\prt\xi^{*n}}
\frac{\prt^l}{\prt(\Omega T)^l}\cdot (-2)^l(\hh/2)^{n+l}
\exp\big[-|\alpha|^2-\gamma T\int_0^t d\tau \bar{I}_{cl}
(\tau,\alpha,\alpha^*) \big]\Big|_{\xi=0,\xi^*=0} \, .
\ee 
The validity of Eq. (\ref{s4_17}), which is the zero approximation,
is the neglecting of the all terms  higher than zero order in Taylor's 
series of  Eq. (\ref{s3_15}). In the classical case, when $\hh=0$, the 
action $\bar{I}_{cl}(\tau,\alpha,\alpha^*)=I_{cl}(\tau,\alpha,\alpha^*)$
for $\xi=\xi^*=0$ corresponds to a chaotic attractor
under condition of strong chaos. A rough estimation of the classical 
chaotic attractor can be obtained from the stability condition {\em e.g.} 
of a limiting cycle with the minimum phase space volume of the order of 
$\ep^2$. In this case, the criterion of the chaotic attractor is 
\be\label{s4_20}
K=4\mu\ep^2 T>e^{\gamma T}>1 \, .
\ee
In the case, when a  chaotic attractor exists, the strongest contribution
to the sums in Eq. (\ref{s3_15}) is due to the derivatives over the 
initial conditions determined by the following term
\be\label{s4_21}
D(I_{cl},I_{cl})=\frac{1}{\hh}\left(\frac{\prt
I_{cl}(\tau,\alpha,\alpha^*)}{\prt\alpha}\right) \cdot
\left(\frac{\prt I_{cl}(\tau,\alpha,\alpha^*)}{\prt\alpha^*}\right)
\propto e^{(\Lambda -\gamma T)\tau} \, .
\ee
Therefore, the validity condition of the performed approximation is
$ D(I_{cl},I_{cl})\sim \exp[(\Lambda-\gamma T) t]< I_{cl}/\hh $.
Finely, we obtain the validity condition on time, which is
a breaking time of the crossover between classical and quantum dynamics 
for the nonlinear kicked oscillator in the presence of dissipation:
\be\label{s4_23}
\tau_{\hbar}^{(d)}=\ln(I_{cl}/\hh)/(\Lambda-\gamma T) \, .
\ee
It worth to underline that this result coincides with the breaking time 
obtained for a quantum chaotic attractor of a web map \cite{izpre03},
where an existence of the classical
chaotic attractor, namely, ``the dying attractor'' 
($\Lambda -\gamma T\rightarrow 0+$) leads to the essential
increasing of the applicability of the semiclassical approximation.
Since the $S$-matrix corresponds to the quantum map of Eq. (\ref{s4_4}) 
considered on the period $T$, 
therefore, the conditions of the $S$--matrix approach application  
for studying the quantum chaotic attractor depends on 
the correlations between the Ehrenfest time $\thh^{(d)}$, the period $T$ 
and the relaxation time $1/\gamma$.
There exist two important inequalities, that are important for the present 
analysis. The first one  is $ T>1/\gamma$, and it means that the 
relaxation process is fast and should not be taken into the
consideration.
The second inequality
$\thh^{(d)}>T$ means that the semiclassical approximation is valid for the
$S$--matrix construction.  In particular, the Ericson 
distribution of fluctuations of the $S$-matrix (\ref{s4_18}) and
(\ref{s4_19}) is valid.
In the same time, a possible phase space 
structure of the chaotic attractor as a cantor set might be studied in 
the framework of the semiclassical $S$-matrix consideration on the finite 
time scale restricted by the breaking time (\ref{s4_23}).

To complete the consideration, the criterion of chaos (\ref{s4_20})
should be explained. By denoting 
$\alpha(\tau)=\sqrt{I(\tau)}e^{-i\theta(\tau)}$ (see also Eqs. 
(\ref{s3_6}) and (\ref{s3_7})),
an approximated map  for the action--angle $(I,\phi)$ variables
has the form
\be\label{s4_24}
I_{\tau+1}=e^{-\gamma T}
[I_{\tau}+2\ep\sqrt{I_{\tau}}\sin\theta_{\tau}+\ep^2] \, , ~~
\theta_{\tau+1}=\theta_{\tau}+\Omega T+2\mu I_{\tau+1}.
\ee
In spite of the map is an approximation of the exact analysis
of \cite{bz}, it is detailed enough in order to obtain a crude
criterion of chaos.
Therefore we obtain for the local instability condition
$|\prt\theta_{\tau+1}/\prt\theta_{\tau}-1|\sim 4\mu\ep T\sqrt{I_{\tau}}
e^{-\gamma t}=K\, e^{-\gamma t}$
The minimum action and correspondingly phase space volume in any 
moment $\tau$ is 
$
min(I_{\tau})=\inf\big[\int_0^{\tau}d\tau'I(\tau')\big]\sim \ep^2 
$.
Therefore the condition for chaotic attractor is $ K\ge 4\mu\ep^2 T 
e^{-\gamma T}>1 $. It also follows that the condition for 
quantum chaos \cite{bz82} $\hh=\hbar \mu T \ll 1,~
\kappa=4\ep^2/\hbar\gg 1$  reads now $\hh\kappa=K >e^{\gamma T}$.

I thank P. Gaspard and I. Guarneri for useful discussions. This work was 
supported by the  Minerva Center for Nonlinear Physics of Complex Systems.

\end{document}